\documentclass[journal=jacsat,manuscript=article]{achemso}
\usepackage[version=3]{mhchem} % Formula subscripts using \ce{}
\usepackage[T1]{fontenc} % Use modern font encodings
\usepackage[usenames,dvipsnames]{color}
\usepackage{hyperref}

\author{Ashutosh Shukla}
\email{ashutosh.shukla@students.iiserpune.ac.in}
\author{Sneha Boby}
\author{Rahul Chand}
\altaffiliation{Present Address: Faculty of Engineering and Natural Sciences, Tampere University, FI-33014, Tampere, Finland}
\author{G. V. Pavan Kumar}
\email{pavan@iiserpune.ac.in}
\affiliation[Unknown University]
{Department of Physics, Indian Institute of Science Education and Research Pune, Pune, Maharashtra, 411008, India}
\title{Rotational Jamming of Plasmonic Optical Matter Driven by Chiral Light}


\begin{document}
\begin{abstract}
Plasmonic Optical matter (OM), composed of optically bound metallic particles, can be rotated by transferring the spin angular momentum (SAM) of chiral light to the assembly. Rotating OM is a promising platform for optical micromachines, with potential applications in plasmofluidics and soft robotics. Understanding the dynamic states of such Brownian, micro-mechanical systems is a relevant issue. One key problem is understanding kinetic jamming and clogging. Studies of driven multiparticle systems have revealed that under suboptimal driving, the systems can stop moving, showing jamming transitions. It is important to identify dynamic regimes where crowding competes with driving and is susceptible to jamming in the context of optical micromachines. Through experiments supported by numerical simulations, we reveal assemblies with well-defined hexagonal or triangular symmetry that efficiently harness the SAM of incident chiral light, resulting in stable rotation. However, as the plasmonic-particle assembly grows and its dimensions approach the beam waist, new particles can disrupt this order. This causes a transition to a ‘fluid-like’ state with less-defined symmetry, correlated with a significant reduction in transferred torque, causing rotation to stagnate or cease. We suggest this behaviour is analogous to a rotational jamming transition, where the rotational motion is arrested. Our findings establish a clear relationship between the structural symmetry of the OM assembly and its ability to harness SAM, providing new insights into controlling chiral light-matter interactions and offering a novel platform for studying jamming transitions.
\end{abstract}

\maketitle

%\tableofcontents

\section{Introduction}
Optical matter (OM) is a dynamic, non-equilibrium phase of matter composed of particles that self-assemble and organize under continuous illumination from an off-resonant laser.\cite{thirunamachandran_intermolecular_1980,burns_optical_1989,mcgloin_optically_2004,mohanty_optical_2004,grzegorczyk_stable_2006,karasek_analysis_2006,ahlawat_optical_2007,karasek_long-range_2008,demergis_ultrastrong_2012,arita_optical_2018} The particles are held together by optical binding force, which is a laser scattering-induced interaction.\cite{dholakia_colloquium_2010,forbes_optical_2020,rodriguez_optical_2008,romero_electrodynamic_2008,demergis_ultrastrong_2012,kudo_optical_2016,kudo_single_2018,wang_optically_2016} This interaction creates stable, dynamic configurations in which particles are held together without physical contact, often with interparticle distances that are integer multiples of the laser wavelength.\cite{yan_guiding_2013,simpson_optical_2017,chen_negative_2014,qi_stable_2022,yan_potential_2014,huang_primeval_2022,mirzaei-ghormishNonlinearOpticalBinding2025b} Optical matter, therefore, represents a unique system for studying self-organization\cite{han_crossover_2018,han_phase_2020,louis_unconventional_2024,tao_generalized_2024,tao_rotational_2023} and collective behavior in driven, non-equilibrium systems. Beyond linear momentum, light also possesses spin angular momentum (SAM), which is associated with its polarization state.\cite{sharma_spin-hall_2018,sharma_optical_2019,paul_focused_2021,paul_simultaneous_2022,bliokh_spinorbit_2015,liaw_spin_2018,kotlyar_spin-orbit_2020,kumar_when_2023} This SAM can be transferred to colloids as well as optical matter assembly, causing them to rotate.\cite{stilgoe_controlled_2022,chang_optical_1998,padgett_light_2000,bruce_initiating_2021,friese_optically_2001,toftul_optical_2025,paterson_controlled_2001,adachi_orbital_2007,zhao_direct_2009,lehmuskero_ultrafast_2013,yu_rotation_2015,shao_light-driven_2018,reichert_hydrodynamic_2004} The magnitude and direction of the imparted torque are determined by the specific structure of the particle assembly and the chirality of the illuminating laser beam.\cite{rodriguez_optical_2008-1,brzobohaty_experimental_2010,sukhov_actio_2015,haefner_conservative_2009,han_crossover_2018,genet_chiral_2022} In assemblies with dynamical states having well-defined rotational symmetry, the scattering of light is predictable. This facilitates an efficient and continuous transfer of angular momentum. The potential application of OM as micromachines, including micro-gears and micro-engines,\cite{figliozzi_driven_2017,parker_optical_2020,wang_microscopic_2025} and in soft robotics relies on the ability to precisely control these dynamical states. 

Understanding the degrees of freedom of such light-driven metallic micromachines has emerged as a relevant problem in optically bound matter. To reveal the physical conditions under which OM can be harnessed as rotational micromachines, it is necessary to interrogate them as micro-mechanical systems subject to Brownian noise. This framework of driven systems with Brownian noise motivates a statistical mechanics viewpoint, where several concepts related to dynamic states have been discussed and studied at micron and sub-micron scales.\cite{nagel_experimental_2017, martinezBrownianCarnotEngine2016} Among the many, jamming of soft matter systems\cite{liu_jamming_1998} has turned out to be of interest not only as a curious observation, but also with technological implications. Jamming transitions are a universal kinetic arrest phenomenon that has been widely studied in macroscopic granular media, foams, and colloids.\cite{chen_optical_2024,sellerio_glass_2011,kumar_structural_2004,zhang_jamming_2005,ravazzano_unjamming_2020,liu_local_2023,zhu_variable_2024,behringer_physics_2018,ganapathi_structure_2021,bi_jamming_2011,henkes_jamming_2005,hima_nagamanasa_direct_2015,zottl_emergent_2016,tan_odd_2022,zaccone_complete_2025} In these systems, increasing particle density leads to a disordered, jammed state where motion is arrested. The efficient driving of the systems requires the identification of a suitable parameter space of particle density and driving.\cite{cereceda-lopez_hydrodynamic_2021} 

In the context of optical matter and its connection to kinetic jamming, several questions arise, such as: Can we observe jamming effects in translational and rotational dynamic states of optically bound matter? What are the conditions under which jamming can be observed and reversed? How can the spatial and temporal aspects of the angular momentum of light influence the jamming and unjamming transition in OM? Motivated by these questions, this paper reports the observation of jamming of rotation in colloidal optical matter driven by the spin angular momentum of light. We specifically identify the conditions under which jamming occurs for a few-particle system and reveal the connection between the symmetry of the assembly and the torque generated in a focused, chiral, spinning optical trap.
\begin{figure*}
\centering
\includegraphics[width= \textwidth]{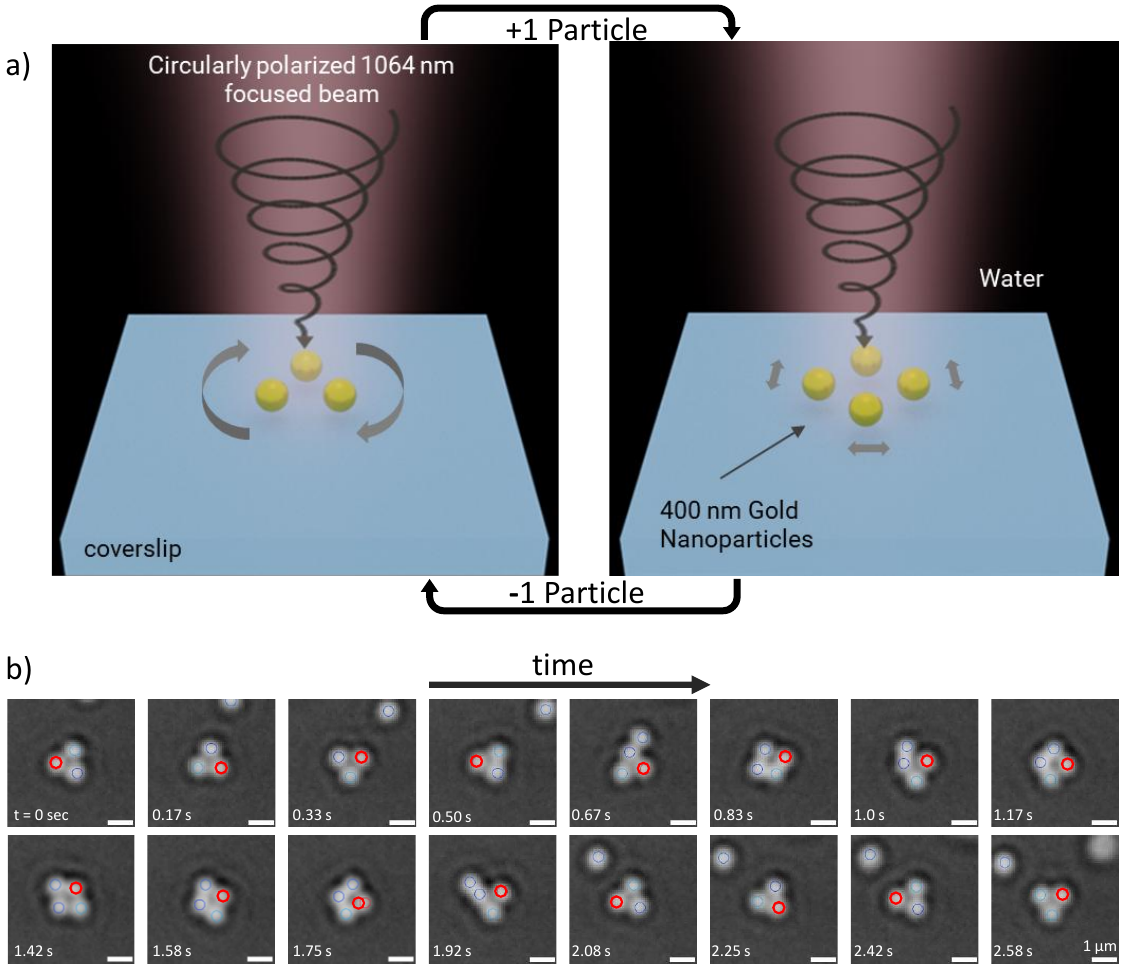}
\caption{Arresting the rotation of the optically bound assembly. a) A schematic figure shows that when circularly polarised light is used to form and rotate Optical matter composed of 3 particles, it rotates stably. Additionally, the assembly has a triangular structure. However, with the addition of one more particle, the assembly loses its triangular symmetry and becomes a fluctuating square. The four-particle assembly also stops rotating. b) The time series of experimental data shows the rotation of the three particle assembly until the addition of a fourth particle. One of the particles is marked by a red circle to improve visualisation. After the fourth particle leaves the assembly due to thermal noise, the assembly starts rotating again.}
\label{schema}
\end{figure*}
In this report, we experimentally and computationally demonstrate the jamming of rotating OM formed by 400 nm gold nanoparticles, assembled and driven by a SAM-carrying focused beam. Focused laser beams provide a more efficient method for optical manipulation,\cite{jones_optical_2015,volpe_roadmap_2023,chand_optothermal_2025,shukla_synchronized_2025,tsujiThermoosmoticSlipFlows2023} and owing to their finite extent, a crowding effect is possible where multiple particles are present in a confined region. It is already known that the behavior of optical matter changes fundamentally under focused beams due to the strong intensity gradients.\cite{yan_potential_2014,huang_primeval_2022} Thus, focused beams provide an excellent platform to study jamming in rotating optical matter. The jamming we observe occurs due to a structural transition, as shown schematically in Figure \ref{schema}a. As a stable, hexagonally packed assembly of gold nanoparticles grows, the addition of a new particle can disrupt its order, causing a dramatic transition from a rigid, crystalline-like structure to a disordered, `fluid-like' state. This loss of symmetry is directly correlated with a significant reduction in transferred torque, causing rotation of the assembly to stagnate or cease. The observations reported here are similar to jamming transitions. Our system, driven by a continuous optical force rather than thermal motion, offers a unique mesoscopic platform for studying these phenomena.

We show, in Figure \ref{schema}b, experimental images of the rotation of the optically bound assembly of 400 nm gold nanoparticles confined with a 1064 nm laser. While there are three particles in the assembly, the assembly rotates stably, as can be observed by noting the position of the particle tagged with a red boundary. As soon as the fourth particle joins this assembly, the rotation is arrested, and the assembly shows stochastic movement. The arrest of the angular momentum transfer is reversible, as the rotation continues as soon as the fourth particle leaves the assembly. Our simulations corroborate this finding, revealing that positive torque is efficiently generated in hexagonally or triangularly packed states, whereas states with other packing arrangements, such as the intermittent square-packed states observed experimentally, result in weak, near-zero, or even negative optical torque. The rotation resumes when the defect particle leaves the assembly and the hexagonal symmetry is restored, establishing a clear link between the structural order of the assembly and its ability to convert the light's spin angular momentum into its own orbital angular momentum.
\section{Results}
Building on the discussed framework, we now present the experimental and computational evidence for the rotational arrest phenomenon. Optically bound nanoparticles have been studied under both focused and broad illumination. Broad illumination requires larger input powers to have sufficient intensity in the sample plane, as it depends inversely on the square of the beam waist. Foundational insights have been gained from OM rotation experiments and simulations, considering both broad-beam and plane-wave illuminations that carry spin angular momentum. But the behaviour of OM fundamentally changes under focused beams due to strong intensity gradients, which is the focus of our study. Thus, we demonstrate the evolution of rotating optically bound matter (OM) under conditions of a circularly polarized focused beam, which enables the creation of assemblies using a minimal power of approximately 20 mW. We illustrate the changes due to focusing by experimenting with two different numerical aperture objectives, which make beam spots of 1.5 \(\mu m\) and 2 \(\mu m\), respectively. We have discussed the theoretical differences in optical binding in plane wave vs focused beam illumination in Supplementary Information S1 and S2\cite{supple}.
\subsection{Arrested Rotation of Four Particle Assembly}
We assembled 400 nm gold nanoparticles (AuNPs) using a 1.2 NA water immersion objective with a 1064 nm laser with 20 mW power in the sample plane. The experiments were performed in an upright configuration, where the laser illuminated the sample chamber from above. The particles are thus pushed down to the glass substrate, where they are confined in two dimensions. The focused beam creates a small beam spot (1.5 \(\mu m\)), which implies a limited number of particles can be accommodated under the laser. The laser is passed through a combination of a polarizer, a half-waveplate, and a quarter-waveplate before reaching the objective lens to make the light circularly polarized or chiral. This circular polarization is manifest as the spin angular momentum of light and leads to the rotation of the assembly. Supplementary Video 1 shows the behavior of 3 and 4 particles under such an illumination at a 3x slowed speed. The video is tracked, showing individual particles, each marked with a differently colored circle. The scale bar in the video is 1 micron. 
\begin{figure*}
\centering
\includegraphics[width= \textwidth]{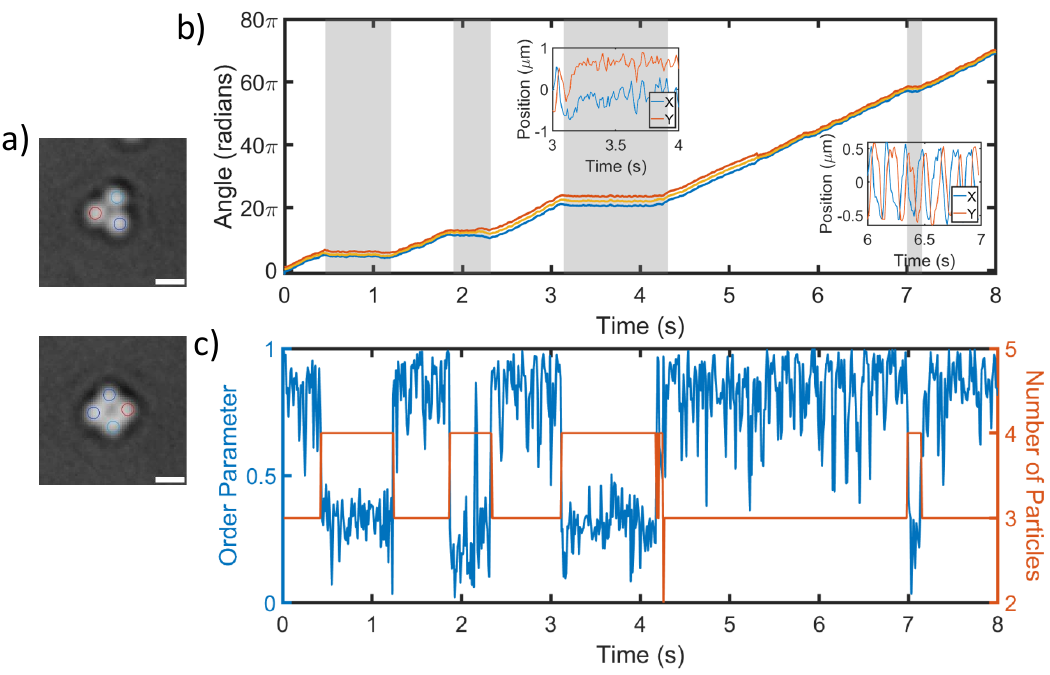}
\caption{Evolution of optical matter from 3 to 4 particles. a) Snapshots of three-particle (3P) and four-particle (4P) assemblies corresponding to the supplementary video 1. This assembly is created with a 1.2 NA objective, creating a tighter spot. The assembly fluctuates from three to four particles intermittently. b) Angular trajectories of particles are shown, showing rotation by an increase in the angles. The 4P assembly times are highlighted in grey. The rotation stagnates whenever the 4th particle joins the assembly, distorting the triangular symmetry. c) An order parameter that quantifies the triangular nature of the assembly is plotted in blue for the whole time. It can be seen that the order parameter drops sharply as the fourth particle joins the assembly. The red line with the y-axis on the right side shows the instantaneous number of particles in the assembly.}
\label{3to4}
\end{figure*}
It shows that the optically bound assembly of three particles, which has a triangular structure, rotates stably. However, when the fourth particle joins the assembly, it distorts the triangular symmetry and also causes a loss of rotation. Figure \ref{3to4}a shows snapshots of the three and four-particle assemblies. Figure \ref{3to4}b shows the angular position of all three particles with time. The times when a fourth particle joins the assembly are marked with a grey highlight. The angular positions stop increasing immediately as the fourth particle joins and distorts the triangular symmetry. The effect is reversible as the particles start rotating again when one of the four particles leaves the assembly.

To correlate the structure of the assembly with the rotation, we defined and calculated an order parameter \(\psi_3=\frac{1}{3} \sum_{i=1}^N \exp(3i\theta_i)\) which quantifies the threefold symmetry of the assembly. Here, \(\theta_i\) is the angle of a particle from the x-axis with the beam centre as origin. We highlight in Figure \ref{3to4}c that the number of particles in the assembly correlates with the magnitude of this (Triangular) Order Parameter. These quantities also correlate with the stagnation of rotation, as can be seen from Figure \ref{3to4}b.

The observed stopping of rotation in our optical matter assembly can be described as a dynamic arrest phenomenon observed in soft matter.\cite{chen_optical_2024,nagel_experimental_2017,eberle_dynamical_2011,reichhardt_active_2014} In this system, the optical torque, which provides the drive or activity, competes with increasing particle number, which introduces a crowding effect. This transition from a rotating, ordered state to a jammed, non-rotating state shares striking parallels with dense active systems, where a transition occurs as the drive competes with the increasing density of active particles. In our case, the addition of a single particle acts as a defect that disrupts the collective symmetry, effectively jamming the system and arresting the motion. This suggests that the interplay between symmetry and torque in light-driven systems can be viewed through the lens of collective dynamics, offering a new perspective on controlling light-matter interactions at the nanoscale.

To understand the effect of beam size and the generalizability of the jamming phenomenon, we next performed a similar experiment with a different objective lens with a lower numerical aperture. This larger beam spot is accompanied by an overall change in the illumination profile, allowing it to accommodate more particles. The intensity gradient also changes, affecting the relative forces on the particles. Despite these changes, we observe a similar loss of rotation in optical matter under circularly polarized Gaussian beams as the size of the assembly increases.
\subsection{Arrested Rotation of Eight Particle Assembly: Effect of Beam Size}
\begin{figure*}
\centering
\includegraphics[width= \textwidth]{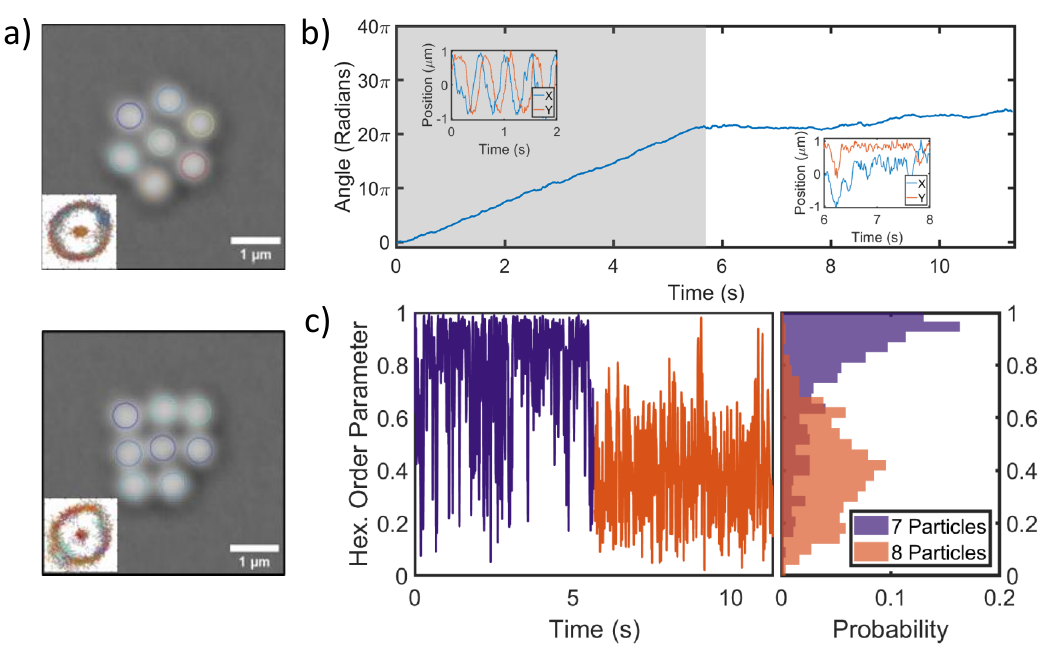}
\caption{Evolution of optical matter from 7 to 8 particles. a) Snapshots of seven-particle (7P) and eight-particle (8P) assemblies corresponding to the supplementary video 2. b) Mean angular rotation of the assembly plotted as a function of time. The transition boundary is marked with a change in colour from grey to white. The rotation of the assembly ceases dramatically upon the addition of the 8th particle. Insets show the position of one peripheral particle in the assembly in Cartesian coordinates for both cases. This again illustrates the rotation before the transition and the stochastic motion that follows. c) The instantaneous local hexagonal order parameter (HOP) of the assembly is plotted for the whole duration of the video. The assembly transitions from 7P to 8P at 5.7 seconds. The 7P assembly is comparatively more hexagonally ordered than the 8P assembly. The distribution of the HOP for both halves is plotted in the adjacent plot.}
\label{7to8}
\end{figure*}
To understand the effect of beam size on our observations, we conducted a similar experiment using a 100\(\times\) objective with a 0.95 NA, which creates a slightly larger beam spot (\(\sim 2\ \mu m\)). This larger spot accommodates more particles. In this case, we observe a similar transition from rotation to non-rotation when the assembly transitions from seven to eight particles (Figure 3a), in contrast to the transition from three to four particles observed with the tighter beam.

This seven-to-eight particle transition is shown in Supplementary Video 2 at a 4 times slower speed. This video is also tracked with individual particles marked with differently colored circles. The 45-second video shows, in the first half, the structure and rotation of 7 particles, followed by the addition of the 8th particle in the 23rd second. The assembly, thereafter, becomes disordered, displaying intermittent states with various packing symmetries. Along with the loss of hexagonal symmetry, the assembly also ceases to rotate. This loss of symmetry and the associated cessation of rotation are reversible. When the 8th particle leaves the assembly, the assembly stabilises and rotates again. The measured mean interparticle separation in this case is nearly 860 nm, and thus the assembly spans a bigger area.

We show representative snapshots of 7P and 8P assembly in Figure \ref{7to8}a. The overall trajectory of all particles is shown in the insets. To quantify the loss of rotation, the mean angular position of the peripheral particles is shown in Figure \ref{7to8}b. The angles increase until the 8th particle joins the assembly, and then stagnate and fluctuate. Finally, to quantify the structural order, we calculate a local hexagonal order parameter.\cite{sharma_large-scale_2020,han_phase_2020} This order parameter is defined as: 
\begin{equation}
\psi_6=\frac{1}{N} \sum_{i=1}^N \exp(6i\theta_i)\label{eq1}
\end{equation} 
where i sums over all the N particles which are neighbours of the central particle in the assembly. Since our system only consists of 7 to 8 particles, the \(\psi_6\) gives information about the sixfold symmetry of the assembly. When the assembly is sixfold symmetric, HOP has a value equal to 1, and decreasing sixfold symmetry implies a decrease in HOP up to 0. This local hexagonal order parameter is plotted for all the frames of the Supplementary Video 2 in Figure \ref{7to8}c. The order parameters for 7P and 8P times are colored differently for contrast. The order parameter reduces for the 8P assembly. We plot the order parameter distribution histogram in the same figure. The histogram shows that for 7P assembly, the HOP is more skewed to 1, which implies a hexagonal packing, whereas the 8P assembly has a peak near 0.4. Typically, particles in a fluid have a HOP value around 0.4.\cite{han_phase_2020} Thus, we refer to this state as a `fluid-like' state.

As noted previously, this loss of rotation is akin to more widely observed arresting phenomena. But in the case of eight particles, we can see the role of a hexagonal order more clearly compared to the four-particle case discussed in the previous section. This transition from an ordered, crystalline-like state (\(\psi_6\approx 1\)) to a disordered, amorphous state (\(\psi_6\approx 0.4\)) can be described as a jamming transition.\cite{libal_colloidal_2013,senbil_observation_2019,korda_evolution_2002,cao_numerical_2003} Our observation represents a rotational jamming transition, a form of universal kinetic arrest phenomenon where the rotational motion of the system ceases.\cite{robertson-anderson_optical_2018,cereceda-lopez_hydrodynamic_2021,lips_hydrodynamic_2022,cereceda-lopez_excluded_2024,saito_change_2018,yan_rotating_2014,stoop_clogging_2018} In this dynamic, light-driven system, the seven-particle assembly maintains its crystalline-like hexagonal structure, enabling stable rotation. However, the addition of the eighth particle disrupts this order, causing the system to exceed a critical packing threshold. The eight-particle assembly, with its disordered structure, represents a kinetically arrested transient phase. This state is unable to efficiently convert light's spin angular momentum into collective orbital motion, causing the rotational drive to be overcome by the loss of symmetry.

To further validate our findings and explore the effect of different parameters, we investigated assemblies of 250 nm gold nanoparticles (AuNPs) using a 785 nm laser, shown in Supplementary Video 3 and discussed in Supplementary Information S3.\cite{supple} While previous literature\cite{qi_stable_2022} has shown that particles ranging from 100 nm to 350 nm can form stable, rotating assemblies, our study demonstrates that 400 nm AuNPs (\(\approx 0.5\lambda_m\)) can also form rotating optical matter. However, the symmetry transition and subsequent rotation arrest are highly dependent on the particle size, laser wavelength, and beam profile. Our experiments confirm that varying these parameters leads to different magnitudes of scattering forces, which in turn give rise to non-hexagonal states, as shown in Figure \ref{evolution}.
\subsection{Connecting the symmetry of the assembly with rotation}
\begin{figure*}
\centering
\includegraphics[width= \textwidth]{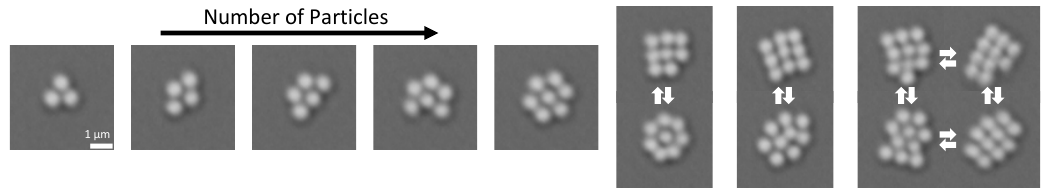}
\caption{Evolution and rotation of optical matter from three to eight particles in focused beams. 400 nm AuNPs are assembled with a 1064 nm laser. Initially, the particles assemble in a hexagonal symmetry up to 7 particles. After seven particles, the hexagonal assembly becomes unstable, and some states begin to emerge with square packing symmetry.}
\label{evolution}
\end{figure*}

The observation of a phase transition from an ordered, hexagonally packed state to a disordered, fluid-like state is a key finding of our experiments, as elaborated in Sections 2.1 and 2.2. This is particularly significant because the hexagonal packing symmetry is a notable aspect of most assemblies shown in the literature. Our results demonstrate that this hexagonal symmetry degenerates into fluid-like states upon reaching a critical size, as shown in Figure \ref{evolution}, which is dependent on the particle size and laser wavelength used. 

We hypothesize that this loss of symmetry is directly responsible for the arrest of rotation. As long as the assembly maintains its symmetry, it will be able to continuously harness the angular momentum from the laser beam. However, for the 400 nm particles, the assemblies enter an amorphous or fluid-like arrested phase, resulting in a chaotic potential energy landscape that lacks a clear, stable energy pathway for rotation. This is in contrast to the ordered, rotating assemblies that are trapped in a well-defined energy minimum. In these non-rotating systems, the assembly exists in a state with multiple, nearly degenerate low-energy configurations, which allows the particles to fluctuate between various positions rather than follow a consistent rotational path. This explains why the rotational drive is overcome by the loss of symmetry, resulting in the kinetic arrest phenomenon observed in Sections 2.1 and 2.2. To confirm this hypothesized link between structure and rotation, we performed electrodynamic force simulations to directly calculate the torque on assemblies with different structural symmetries. The following section details these computational findings and establishes a direct link between the structure of the assembly and its rotational behavior.
\subsection{Static Electromagnetic Force Simulations: Correlating Torque and Symmetry}
\begin{figure*}
\centering
\includegraphics[width=0.8\textwidth]{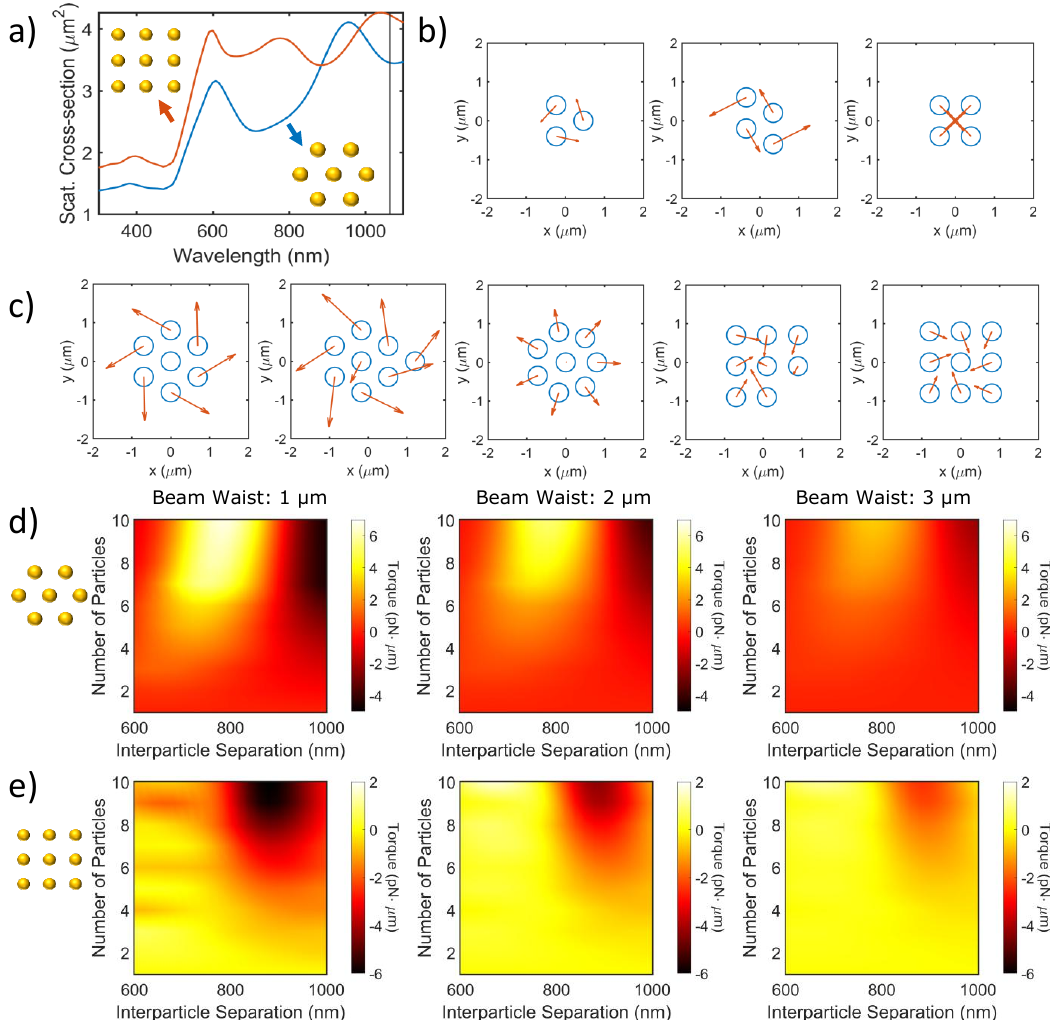}
\caption{Symmetry-dependent electromagnetic interaction and force. a) The scattering cross-section of optical matter assemblies is studied as a function of the assembly symmetry, specifically comparing a hexagonally-packed and a square-packed array. b) The force on particles for 3P and 4P assemblies with different symmetries from experimental data is calculated using GMMT-based simulations. c) The force on particles for 7P and 8P assemblies. The particle arrangement corresponds to representative positions with different symmetries as shown in Figure \ref{evolution}b. d) and e) show the total torque on the assemblies as a heatmap for different numbers of particles with hexagonal and square symmetry, respectively. The torque is calculated for different particle separations. The torque for hexagonal arrays with 800 nm separation increases with the number of particles. Meanwhile, the torques are mostly small and negative for a square array.}
\label{Torque}
\end{figure*}

Our experimental results demonstrate a clear correlation between the loss of structural symmetry and the cessation of rotation. To computationally validate this finding, we used a generalized multiparticle Mie-theory-based Python package, MiePy\cite{parker_optical_2020}, to perform electrodynamic simulations and calculate the optical forces and torques on static assemblies. We focus on the crucial difference in torque generated by two key structural symmetries: the hexagonal packing observed in our stable, rotating assemblies and the square packing that appears as an intermittent, disordered state in our experiments. We can observe that the scattering cross sections of a square and hexagonal array shown in Figure \ref{Torque}a exhibit a clear difference. Thus, trivially, the interaction of these OM arrays with light is significantly different and dependent on the structure.

As established in previous studies, the transfer of torque to an assembly is fundamentally linked to its structural symmetry.\cite{chen_negative_2014,han_crossover_2018,tao_rotational_2023} When a photon with an initial angular momentum, \(m_i\), scatters from a structure possessing a discrete \(m_s\)-fold rotational symmetry, the final angular momentum, \(m\), of the scattered photon is constrained to specific allowed channels. The relationship is described by the equation:\[m=m_i+n\times m_s,\] where n is an integer \((n=0,\pm1,\pm2,...)\). Each of these scattering channels has a distinct cross section, meaning that the probability of a photon scattering into a particular channel varies.\cite{tao_rotational_2023,qi_stable_2022} The difference in the final and initial angular momenta of the photons is taken by the assembly, thus forming rotating optical matter.

In assemblies with well-defined symmetries, the scattering modes of the scattered photons are also well defined and stable. This predictable scattering behaviour facilitates an efficient and continuous transfer of SAM from the incoming light to the assembly, which is then converted to the OAM of the rotating particle cluster. This process results in a consistent and sustained rotation of the optical matter. In contrast, for assemblies that lack a stable, well-defined symmetry, such as the `fluid-like' states observed in experiments, the scattering modes constantly fluctuate as the particles move and reconfigure. This continuous change in the scattering landscape disrupts the efficient transfer of angular momentum, resulting in weak or nonexistent torque on the assembly and, consequently, the cessation of rotation.

Thus, based on our experiments, we computationally investigated the transfer of torque to OM arrays under conditions of a circularly polarized, focused beam. See methods and Supplementary Information S4\cite{supple} for further simulation details. We chose square symmetry for comparison, since many of the intermittent states in the fluid-like state of the assembly have square packing. This is visible in Supplementary Videos 3 and 4, as well as in the snapshots of assemblies in Figures \ref{evolution} and \ref{3to4}a.\\
Our simulations validate our experimental observations. For a 3P/4P system, Figure \ref{Torque}b shows the force felt by the particles in a quiver plot. The length of the force arrows is proportional to the magnitude of the force for the three panels, allowing for easy comparison. We follow the convention that the laser hits the particles from above and enters the plane. We use a right circularly polarised light, which produces a counterclockwise torque on the particles. The three particles kept in a triangular configuration are acted on by forces that tend to move them in a counterclockwise manner. The magnitude of total torque on the assembly is 0.768 \(pN\cdot\mu m\). The four particles arranged in a triangular symmetry also show similar forces pointing counterclockwise consecutively with a total torque of 2.018 \(pN\cdot\mu m\). However, the 4P arranged in a square has a vanishing torque of 0.065 \(pN\cdot\mu m\) as the force points towards the centre. 

We can also look at the force on individual particles in a 7P and 8P array with different symmetries. Figure \ref{Torque}c shows forces on the seven and eight 400 nm AuNPs arranged corresponding to Figure \ref{7to8}b in another set of quiver plots. The length of the force arrows is again proportional to the magnitude of the force for the five panels, allowing for easy comparison. The 7P assembly, having hexagonal symmetry, has forces that lead to a torque of magnitude 4.7 \(pN\cdot\mu m\). Consequently, the array should rotate counterclockwise. The 8P hexagonally packed assembly also shows counterclockwise-pointing forces and has a total torque of 6.0 \(pN\cdot\mu m\). Next, we arranged the eight particles with one particle at the centre and the remaining particles in a circle at an 800 nm radius. This assembly corresponds to one of the 8P arrangements in Figure \ref{evolution}b. This assembly shows a small force pointing approximately radially outwards on all particles. This produces a small torque of -0.17 \(pN\cdot\mu m\). This torque is small and would lead to a clockwise rotation. Similarly, the 8P square assembly not only has smaller force magnitudes but also forces that point in the opposite direction, i.e., clockwise, exhibiting a weak negative optical torque of -1.37 \(pN\cdot\mu m\). We also demonstrate torque on a 9-particle assembly, forming a complete 3x3 square. This also shows the negative torque of magnitude -1.4 \(pN\cdot\mu m\). 

We calculate the orbital torques from these forces and study these torques as the static OM assembly evolves from a single particle to multiple particles. We show in Figures \ref{Torque}d and e the total orbital torque as heatmaps on OM assemblies for different numbers of particles with hexagonal and square packing symmetry, respectively. When a 1064 nm laser is used to create optical matter, the optical binding distance is nearly \(\frac{1064}{1.33} \approx 800\) nm. However, since the particles are subject to fluctuations in a fluid, we look at the torque for different lattice spacings. We also studied the torque for these two cases to change the beam waist of the focused beam. The primary effect of increasing the beam waist is that the torque magnitude reduces. The negative torque regime is also shifted to higher interparticle separations. For the hexagonal packing with a lattice spacing of 800 nm, the torque values increase positively with the number of particles. Meanwhile, for square lattice spacings, it is near zero and even negative. For a beam waist of 1 \(\mu m\), the heat map shows a band of low and negative torque for four particles as a function of interparticle separation. This shows a strong dependence of the torque transferred on the structure of the OM. For the square-packed assembly, the torque values are predominantly negative and small for lattice spacing of 800 and 1000 nm. For a lattice spacing of 600 nm, the torque is positive but smaller by an order of magnitude compared to the hexagonal assembly. The torque study on static assemblies establishes a dependence of the torque on the structure. Particularly, a sufficient positive torque is only generated in triangular assemblies with a lattice spacing of 800 nm, which is the theoretical optical binding distance given by \(d_{OB}=\lambda_m=1064/1.33\). And a negative or small positive torque for square assembly with the same parameters. We performed similar simulations to study the behaviour of assemblies formed using different laser beams of gold nanoparticles of different particle sizes. The results are presented in Supplementary Information S5.\cite{supple}
\begin{figure*}
\centering
\includegraphics[width=0.8\textwidth]{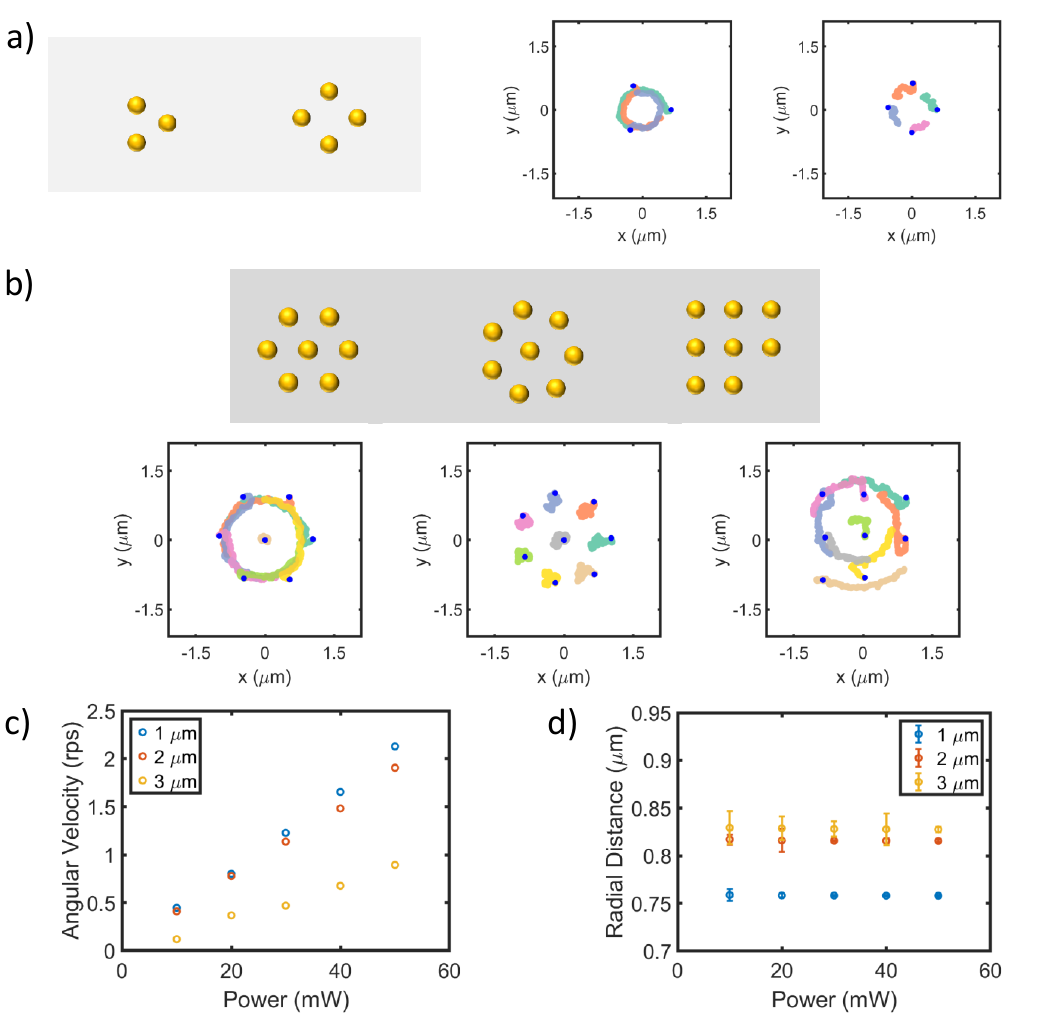}
\caption{Electrodynamic-Langevin dynamic simulations. a) Initial positions for 3P and 4P assembly and the tracks generated by the dynamic simulation. The 3P assembly rotates at a faster rate than the 4P assembly. Additionally, the 4P assembly initially arranged as a square degenerates to a parallelogram with triangular symmetry. The simulations are shown in supplementary video 3. b) Initial positions for 7P and 8P assemblies and the tracks generated by the dynamic simulations. The eight particles are arranged in two ideal representative positions, viz., in a circle and a square arrangement. The 7P arrangement rotates stably. One particle track is marked in purple for visual clarity. The circularly arranged 8P assembly does not rotate around the beam. The square 8P array degenerates and rotates slowly. The simulations are shown in supplementary video 4. c) The 7P dynamic simulation with varying beam widths. The reduced intensity leads to a lower angular velocity. d) The stable optical binding distance is shown for differing beam widths as a function of power.} 
\label{EDLD}
\end{figure*}
\subsection{Electrodynamic-Langevin dynamic simulations show the dynamical behaviour}
While the static force calculations strongly correlate structure with torque, a dynamic model is necessary to capture the time evolution and instability observed in the experiments. We achieve this by solving the overdamped Langevin equation (shown below), which incorporates the optical forces and torques calculated in the previous section.
\begin{equation}
\gamma_i\dot{\vec{r_i}}=\overrightarrow{F}^O_i+\sum_{j\ne i}\overrightarrow{LJ}_ij +\sqrt{2k_BT_i\gamma_i}\overrightarrow{W_i} 
\end{equation}
where \(\gamma_i\) is the viscous drag coefficient and \(r_i\) is the position of the i-th particle. The forces on the right side are the optical force (\(\overrightarrow{F}^O_i\)), the Lennard-Jones interaction force (\(\overrightarrow{LJ}_ij\)), and the stochastic noise-induced random force representing the Brownian fluctuations. The optical force is calculated using the MiePy package at each time step. This Electrodynamic-Langevin Dynamic (EDLD) simulation approach allows us to track the time evolution of the assemblies with different initial symmetries.

The simulations for the three- and four-particle assemblies, shown in Figure 6a and Supplementary Video 5, provide a dynamic confirmation of our experimental results. The Electrodynamic-Langevin dynamic (EDLD) simulation comparing the triangular 3P assembly and square 4P assembly is shown in Supplementary Video 5. The tracks generated from the simulation are summarised in Figure \ref{EDLD}a. The initial positions for both cases are shown on the left. The tracks generated by the time evolution are shown on the right with initial positions marked with blue dots. Both tracks are for the same duration. The 3P assembly rotates stably, maintaining the triangular formation, but the 4P assembly, which started as a square, does not rotate as much at the same time. The square configuration exhibits instability, and in other simulations, it evolves into a triangular arrangement due to stochastic noise. This triangular arrangement shows rotation thereafter.

We extended these dynamic simulations to larger assemblies by modeling the behavior of 7P and 8P arrays. We set up the initial condition for the 7P simulation by placing one particle at the origin and symmetrically placing the six other particles around in a circle. This is shown schematically in the top panel of Figure \ref{EDLD}b. For the 8P assembly, we take two separate initial positions similar to the states observed in the experiment (see Figure \ref{evolution}b). Firstly, we consider a circular assembly with one particle in the centre and seven particles symmetrically around a circle of radius 0.8 \(\mu m \). Next, we make a square-packed state of 8 particles. The comparison of the evolution for these three states is shown in Supplementary Video 6. The corresponding trajectories are shown in the bottom panel of Figure \ref{EDLD}b. The 7P assembly exhibits stable rotation, and the trajectories overlap. The yellow trajectory is overlaid on the other trajectories for clear visualization. The 8P circular assembly does not show orbital rotation as seen for the 7P assembly. In contrast, this assembly slowly orbits the beam axis rigidly, as shown in the corresponding track. In comparison, the square-packed 8P assembly shows instability and rotates slowly after reordering to a hexagonal state. We conclude from the EDLD simulations that triangular or hexagonal states are stable and efficiently harness angular momentum from light in a positive manner. In contrast, the square-packed states are unstable and degenerate into other non-square-packed states, and do not show rotation.

We focus on the stable rotating state of 7 particles and study its properties as the laser power and beam waists are changed in EDLD simulations. In Figure \ref{EDLD}c, we show the dependence of the angular velocity of the assembly. The angular velocity increases linearly with power, and changing the beam waist changes the slope of the linear increase. This is expected as increasing the beam waist reduces the intensity quadratically (\(I\sim w_0^{-2}\)) in the sample plane. We also observe that the stable orbiting radius of the surrounding particles varies as a function of the beam waist. We show this in Figure \ref{EDLD}d. Generally, reducing the beam waist decreases the stable orbiting radii. This can be explained by the fact that the beam exerts a smaller scattering force on the peripheral particles when the beam waist is smaller. We performed EDLD simulations to investigate the behavior of assemblies composed of particles of varying sizes and materials. The results are presented in Supplementary Information S6.\cite{supple}

In this section, we demonstrated the dynamic evolution of OM assemblies using EDLD simulations. The simulations qualitatively support our experimental results, establishing a connection between the structure of the assembly and the torque transferred to it. While our simulations support the experimental findings, it's important to discuss key discrepancies that provide valuable insights and avenues for future research. Our dynamic simulations of an 8-particle assembly, starting from a hexagonally packed state, show stable rotation. However, our experiments with a focused 1064 nm laser beam rarely produced a stable 8-particle hexagonal assembly. Instead, the assembly typically exhibited a fluid-like behavior. This suggests that the experimental conditions introduce factors not fully captured by our simplified model. The discrepancy could arise from several unaccounted factors. Focused beams have a strong intensity gradient not just in the radial direction but also along the beam axis. As the assembly grows in size and its size approaches the beam waist, the outermost particles may experience axial forces leading to their out-of-plane fluctuations. These out-of-plane movements could be disrupting the optical binding and leading to the observed `fluid-like' state. Our simulations, which solve the motion in two dimensions only, would not capture this behaviour. Another potential factor we have largely ignored in calculating the force and motion is the presence of the glass cover slip and associated surface effects. We do so, as it has been previously shown that the presence of the surface does not affect the optical forces in similar experiments. Finally, the model solves the dynamical problem with perfectly identical spheres. This assumption may not hold true in the laboratory, as nanoparticle samples often exhibit considerable polydispersity, which could affect the precise arrangement and stability of the optical matter assemblies. Despite these model simplifications, the simulations have provided crucial insights by establishing a clear link between structural symmetry and the transfer of torque in focused beams.
\section{Discussion and Conclusion}
In this work, we experimentally demonstrated a jamming transition in the rotation of optical matter driven by a focused, chiral laser beam carrying spin angular momentum. We corroborate our observations computationally and reveal a crucial link between the structural symmetry of optical matter assemblies and their rotational dynamics under focused laser beams. We found that stable, rotating assemblies with a well-defined hexagonal or triangular symmetry can undergo a dramatic transition to a disordered, `fluid-like' state upon the addition of a single defect particle. This structural transition is not merely a change in geometry but a rotational jamming transition, representing a dynamic arrest phenomenon where the system's rotational motion ceases.

We experimentally find that the structure of the assemblies can transition from a hexagonal order to non-hexagonal, fluid-like states as a function of particle and incident beam characteristics. We demonstrate that the assemblies remain stable up to a certain number of particles, depending on the beam width. After a critical size is reached, the addition of another particle, which acts like a defect, leads to geometrically frustrated configurations, creating a chaotic potential energy landscape that lacks a clear rotational pathway. The resulting assembly exists as a kinetically arrested transient phase, fluctuating between multiple low-energy states where the rotational drive is overcome by the loss of symmetry. This study builds on the previously explored OM rotation under broad-beam illumination and also establishes that the behavior of optical matter is fundamentally different and more complex under the strong intensity gradients of a focused beam. We corroborate these experimental results with force calculations on static assemblies showing torque behaviour for assemblies of different symmetries. In particular, we compare triangular and hexagonal packed states for three and seven particles with square packed states for four and eight particles, as well as circularly packed states for eight particles. We also perform electrodynamic-Langevin dynamic simulations, which show that the triangular and hexagonal states rotate stably, while square-packed states are unstable and reorder to a hexagonal state. And the circularly packed eight-particle arrangement does not show any orbital rotation, despite being electrodynamically stable for short times.

This finding provides crucial insights into the fundamental physics of light-matter interactions and offers a new principle for controlling microscopic assemblies. Symmetry, in this context, serves as a direct control value for angular momentum dissipation, enabling us to switch the system between an ordered, rotating state and a disordered, non-rotating state. By demonstrating a link between a chiral light-driven force and universal phenomena like jamming transitions and kinetic arrest, this work establishes a unique, non-thermal platform for exploring these concepts in active systems.

The ability to induce a reversible structural transition and a corresponding shift to a near-zero or negative torque regime also opens up possibilities for designing novel, reconfigurable micromachines. Optical matter assemblies can be viewed as prototypes for dynamic micro-machines, whose function can be precisely controlled by manipulating their number density and, consequently, their symmetry. Our work provides an understanding of the mechanism behind the jamming of driven optical matter. This mechanism could inform the development of sophisticated particle sorters, microfluidic pumps, and optical actuators that are precisely controlled by manipulating the structural properties of the assemblies themselves, paving the way for a new generation of light-driven microscopic technologies.
\section{Methods}
\subsection{Materials}
Milli-Q water was used to prepare all samples. Gold nanoparticles with diameters of 150 nm, 250 nm, and 400 nm were purchased from Sigma-Aldrich. The nanoparticles are stabilised in a citrate buffer. Silver nanoparticles with a diameter of 200 nm, stabilised in citrate buffer, were purchased from NanoComposix. We centrifuge the gold nanoparticles and then redisperse them in a more dilute aqueous solution for use in trapping experiments.

Goldseal Cover Glass, purchased from Ted Pella, Inc., was used after cleaning with acetone. The microfluidic chambers for containing samples for optical trapping were fabricated on the cover glass using a 120 \(\mu \)m thick double-sided adhesive spacer, SecureSeal imaging spacer, from Grace Bio-Laboratories.
\subsection{Optical Measurements}
The experiments were performed on two custom-built microscopes. One is an upright microscope, and the other is a dual-channel microscope. In both cases, the trapping is done in upright geometry. We use objectives with different magnifications and numerical apertures. The laser beams are not expanded before the objective and thus underfill the back aperture. We use a polariser, a half-waveplate, and a quarter-wave plate to effectively control the polarisation of light reaching the objective lens. The light collected from the sample plane is filtered using a suitable filter to remove the trapping laser light.
\subsection{Particle Tracking}
The images captured were tracked using the Trackmate plugin of the FiJi distribution of ImageJ\cite{schindelin_fiji_2012}. The images were taken using brightfield illumination, and the gold nanoparticles are visualised as dark objects on a bright background. We invert these images for tracking using the TrackMate plugin. The images extracted from the tracked processed video show particles as white objects on a black background.
\subsection{Simulations}
We use a Python-based generalised multiparticle Mie theory package, MiePy\cite{parker_optical_2020}, to calculate all the optical forces and scattering cross-sections. We solve the overdamped Langevin equation for various numbers of particles and initial positions in Python. At each time step, the optical force on each particle is calculated using MiePy. The time step for the simulations is 1 \(\mu s\).
\section*{Acknowledgements}
The authors thank Dr Sunny Tiwari for their valuable discussions on this project. AS acknowledges the Ministry of Education, Government of India, for the Prime Minister's Research Fellowship. This work was partially funded by AOARD (grant number FA2386-23-1-4054) and the Swarnajayanti fellowship grant (DST/SJF/PSA-02/2017-18) to G.V.P.K.
\section*{Supporting information}
The following files are available at following \href{https://drive.google.com/drive/folders/1itgABeXCE-EscFC7U1zjBlvk-6wZxDRh?usp=sharing}{link}.
\begin{itemize}
\item Supporting Information pdf: Optical binding force comparison in plane wave vs focused beam illumination, Simulation details: Electrodynamics and Langevin dynamics, Optical binding of two particles, Torque on assemblies of different sizes of AuNP, EDLD simulations for different particle sizes and wavelengths.
\item Supplementary Video 1: Evolution of 250 nm gold nanoparticles optically bound assembly with a 785 nm circularly polarised laser beam.
\item Supplementary Video 2: Evolution of 400 nm gold nanoparticles optically bound assembly with a 1064 nm circularly polarised laser beam.
\item Supplementary Video 3: Evolution of 400 nm AuNP and 1064 nm laser beam (focused with 0.95 NA) optical matter from 7 to 8 particles.
\item Supplementary Video 4: Evolution of 400 nm AuNP and 1064 nm laser beam (focused with 1.2 NA) optical matter from 3 to 4 particles.
\item Supplementary Video 5: Electrodynamic-Langevin dynamic simulations comparing 3 and 4-particle assemblies with triangular and square initial positions.
\item Supplementary Video 6: Electrodynamic-Langevin dynamic simulations comparing 7 and 8-particle assemblies with hexagonal and non-hexagonal initial positions.
\end{itemize}
\bibliography{apssamp}% Produces the bibliography via BibTeX.
\end{document}